\PassOptionsToPackage{unicode}{hyperref}
\PassOptionsToPackage{hyphens}{url}
\documentclass[
]{article}
\usepackage{xcolor}
\usepackage{amsmath,amssymb}
\usepackage{iftex}
\ifPDFTeX
  \usepackage[T1]{fontenc}
  \usepackage[utf8]{inputenc}
  \usepackage{textcomp} 
\else 
  \usepackage{unicode-math} 
  \defaultfontfeatures{Scale=MatchLowercase}
  \defaultfontfeatures[\rmfamily]{Ligatures=TeX,Scale=1}
\fi
\usepackage{lmodern}
\ifPDFTeX\else
\fi
\IfFileExists{upquote.sty}{\usepackage{upquote}}{}
\IfFileExists{microtype.sty}{
  \usepackage[]{microtype}
  \UseMicrotypeSet[protrusion]{basicmath} 
}{}
\makeatletter
\@ifundefined{KOMAClassName}{
  \IfFileExists{parskip.sty}{%
    \usepackage{parskip}
  }{
    \setlength{\parindent}{0pt}
    \setlength{\parskip}{6pt plus 2pt minus 1pt}}
}{
  \KOMAoptions{parskip=half}}
\makeatother
\usepackage{color}
\usepackage{fancyvrb}

\DefineVerbatimEnvironment{Highlighting}{Verbatim}{commandchars=\\\{\}}
\newenvironment{Shaded}{}{}

\newcommand{\DataTypeTok}[1]{\textcolor[rgb]{0.56,0.13,0.00}{#1}}

\newcommand{\ErrorTok}[1]{\textcolor[rgb]{1.00,0.00,0.00}{\textbf{#1}}}

\newcommand{\FunctionTok}[1]{\textcolor[rgb]{0.02,0.16,0.49}{#1}}

\newcommand{\OtherTok}[1]{\textcolor[rgb]{0.00,0.44,0.13}{#1}}

\newcommand{\StringTok}[1]{\textcolor[rgb]{0.25,0.44,0.63}{#1}}

\usepackage{longtable,booktabs,array}
\usepackage{caption}
\usepackage{calc} 
\usepackage{etoolbox}
\makeatletter
\patchcmd\longtable{\par}{\if@noskipsec\mbox{}\fi\par}{}{}
\makeatother
\IfFileExists{footnotehyper.sty}{\usepackage{footnotehyper}}{\usepackage{footnote}}
\makesavenoteenv{longtable}
\usepackage{bookmark}
\IfFileExists{xurl.sty}{\usepackage{xurl}}{} 
\makeatletter
\@ifundefined{xmpquote}{}{}
\makeatother
\hypersetup{
  hidelinks,
  pdfcreator={LaTeX via pandoc}}

\author{}
\date{}

\begin{document}

\section{Never Emitted: Reporter Attribution in GitHub's
Machine-Readable Vulnerability
Records}\label{never-emitted-reporter-attribution-in-githubs-machine-readable-vulnerability-records}

\textbf{Anas Mohiuddin Syed} Independent Researcher, Chicago, IL, USA
ORCID 0009-0005-3736-6430 anasmohiuddinsyed@gmail.com

\subsection{Abstract}\label{abstract}

The CVE record format defines a \texttt{credits} container that names
who found or reported a vulnerability, with a typed role per entry. The
OSV schema defines an equivalent field. GitHub, which assigns CVE
identifiers for advisories in its ecosystems, collects this information
from reporters, requires them to accept it, displays it on the advisory
page, and serves it through its own REST API. It emits it into neither
standardized format. Across 238 GitHub-assigned CVE records whose linked
advisory publicly credits at least one party, zero carry a
\texttt{credits} container, while all 238 carry \texttt{metrics} and
\texttt{problemTypes}, two fields the CVE schema leaves optional exactly
as it leaves \texttt{credits}. Across 302 advisories in the same pool we
retrieved GitHub's own OSV export file, and zero carry a
\texttt{credits} field. The omission is not a property of either format:
the Erlang Ecosystem Foundation populates the CVE field on 18 of 18
records in the same pool using a freely available client for CVE
Services. In a census of all 4,889 published CVE records in a two-week
window, 46.9\% carry credits, GitHub's rate is 0 of 570 without any
advisory filter, and assigner behaviour is concentrated at the extremes
without being exhausted by them: 16 assigners emit the field on no
record and 13 on essentially every record, while 8 assigners covering
23\% of the records sit in between. A request to close the gap has been
open since January 2023; GitHub's stated reason for deferring it is
quoted verbatim. We further show that the NVD API schema defines no
credits field, so attribution that CNAs do emit does not reach the
database most tooling consumes: of 43 credit-bearing records traced from
advisory to CVE record to NVD, none retained it. We release the
collection scripts and a frozen snapshot of every API response.

\textbf{Index terms:} CVE, OSV, NVD, vulnerability disclosure,
attribution, metadata completeness, mining software repositories.

\subsection{1 Introduction}\label{introduction}

A CVE record can name the person who found the vulnerability it
describes. The v5 record format defines a \texttt{credits} container for
this purpose, with a free-text value and a typed role per entry, and
roles including finder, reporter, analyst, coordinator, and remediation
developer {[}1{]}. The OSV schema defines a parallel \texttt{credits}
array with name, contact, and type {[}24{]}. Public credit is the
standard non-monetary compensation in coordinated disclosure {[}15{]},
{[}16{]}, and for researchers without institutional affiliation the
record is often the only durable citable trace of the work. Discovery
provenance is also a triage input: a finding from a vendor's internal
audit and a finding reported by an external party have different
histories.

This paper measures whether those fields are populated, and concentrates
on the case where the answer is unambiguous. GitHub is a CNA and assigns
CVE identifiers for advisories in its ecosystems. Its advisory system
solicits credit from reporters, requires the credited account to accept
before display, assigns a typed role, and serves the result through the
public advisories REST API {[}19{]}. GitHub's own changelog states that
its credit types ``mirror those in the CVE 5.0 schema'' {[}21{]}. The
data therefore exists, in structured form, typed against the target
schema, in the same system that authors the CVE record.

We find it is not emitted into either machine-readable artifact GitHub
produces. Across 238 GitHub-assigned records in our pool, all of which
link an advisory that publicly credits at least one party, the
\texttt{credits} container is absent from every one. The same 238
records carry \texttt{metrics} and \texttt{problemTypes} without
exception, two fields the CVE schema makes optional exactly as it does
\texttt{credits}. The same pattern holds in GitHub's OSV export.

Our contribution is not the discovery of the phenomenon, which
practitioners have discussed since January 2023. It is the
quantification: the rate, its consistency across six independent
observation points spanning 26 months, its extent across both of
GitHub's machine-readable surfaces, the contrast with CNAs operating
under the same format with free tooling, and a population census that
places the practice in context. We ask:

\textbf{RQ1.} Do GitHub-assigned CVE records carry the reporter
attribution that the corresponding advisory displays, and do they carry
other optional CNA-supplied structured fields?

\textbf{RQ2.} Does GitHub's OSV export carry it?

\textbf{RQ3.} How does this compare to other assigning organizations,
both within the same pool and in a census of the record population?

\textbf{RQ4.} Where credits are emitted, do they survive into the NVD?

\subsection{2 Background}\label{background}

The CVE Program assigns identifiers to publicly disclosed
vulnerabilities. Records are authored by CNAs, of which 543 are listed
as of August 26, 2026 {[}26{]}, and served as JSON by CVE Services. The
v5 format {[}1{]} places CNA-supplied content in a
\texttt{containers.cna} object; \texttt{credits} has been present since
version 5.0 {[}2{]}. A \texttt{source} object holds loosely structured
provenance, commonly an originating advisory identifier and a discovery
method. The format is maintained by the CVE Quality Working Group, which
adopted a Request for Discussion process for format changes in 2025
{[}3{]}.

The NVD ingests published records, adds CVSS scores, CWE assignments,
and CPE product identifiers, and republishes through its API 2.0
{[}18{]}. The NVD copy, not CVE Services, is what most scanners and
downstream databases consume, a dependence made visible when NVD
enrichment slowed in 2024 {[}10{]}, {[}11{]}.

The GitHub Advisory Database curates advisories for open source
packages. Reviewed advisories carry credit entries tied to GitHub
accounts with typed roles {[}21{]}. Credited users must accept the
credit before it is displayed. GitHub publishes this database in two
machine-readable forms: the advisories REST API {[}19{]}, which serves
credits, and a repository of OSV-format files {[}25{]}, which is the
bulk export most downstream databases mirror. The reviewed subset is
therefore a sampling frame in which public credit and CVE identifier are
available together, and GitHub is a case where the CNA demonstrably
holds the data its records omit.

Note that the CVE record and the OSV file are distinct artifacts with
distinct schemas. The public request discussed in Section 5 was filed
against the OSV export. We measure both surfaces separately and do not
treat a statement about one as a statement about the other.

\subsection{3 Methodology}\label{methodology}

We construct four datasets.

\textbf{D1, the advisory-linked cluster sample.} We queried the GitHub
advisories REST API {[}19{]} for reviewed advisories that both credit at
least one party and carry a CVE identifier, issuing one query per
half-year window from January 2024 to August 2026 and retrieving up to
three pages per window. The API returns results in descending
publication order, so each window's three pages are drawn from the end
of that window rather than spread across it. The realized sample is
therefore not a tiling of the interval but a cluster sample: six
clusters of 4 to 18 days each, one per half-year, spanning June 17, 2024
to August 26, 2026. We describe it this way because the distinction
matters for what the pool can support, and we report per-cluster results
in Section 4.1 so that consistency across clusters can be judged
directly. This produced 309 advisories. For each we retrieved the
corresponding record from CVE Services and evaluated whether
\texttt{containers.cna.credits} exists and is non-empty, recording the
assigning organization, the \texttt{source} object, the full description
text, every \texttt{containers.adp} structure, and the presence of
\texttt{metrics}, \texttt{problemTypes}, \texttt{affected},
\texttt{references}, and \texttt{descriptions}. 303 records were
retrieved. Six returned HTTP 404: the identifier appears in a published
GitHub advisory but no corresponding record is served by CVE Services.
We report these in Section 4.6 rather than treating them as missing
data.

\textbf{D2, the population census.} To obtain a base rate independent of
the GitHub frame, we enumerated every CVE identifier the NVD lists as
published between June 1 and August 14, 2026, which is 5,037
identifiers, and retrieved the upstream record for each from CVE
Services. This is a census of a fixed window, not a sample, so it
carries no sampling error and requires no seed. It applies no advisory
filter.

\textbf{D3, the linked trace.} D1 contains 43 records that do carry
credits. We queried the NVD for each to determine whether the field
survives republication, and searched the entire returned object for both
a \texttt{credits} key and the literal reporter names present upstream.
This traces the same records across both hops rather than comparing
disjoint samples. NVD applies a rate limit to unauthenticated clients;
we paced requests at 6.5 seconds with exponential backoff on failure,
and record the response category for every identifier rather than
dropping failures.

\textbf{D4, the OSV export.} For every advisory in D1 we retrieved
GitHub's OSV-format file from the advisory database repository {[}25{]}
and recorded whether it contains a \texttt{credits} key.

All responses were retrieved on August 26, 2026 and archived verbatim.
Proportions are reported with exact (Clopper-Pearson) 95\% confidence
intervals, rounded to one decimal place below 10\% and to whole percent
above. Where we compare two assigners we report Fisher's exact test.

\subsection{4 Results}\label{results}

\subsubsection{4.1 RQ1: GitHub emits every other optional field, and no
credits}\label{rq1-github-emits-every-other-optional-field-and-no-credits}

Table 1 reports field presence across the 238 GitHub-assigned records in
D1. Every one of these records links an advisory that publicly names at
least one credited party.

\textbf{Table 1. CNA-container field presence, GitHub-assigned records
(n = 238).}

{\def\LTcaptype{none} 
\begin{longtable}[]{@{}lllll@{}}
\toprule\noalign{}
Field & CVE schema status & Present & Share & 95\% CI \\
\midrule\noalign{}
\endhead
\bottomrule\noalign{}
\endlastfoot
\texttt{affected} & required & 238 & 100\% & {[}98, 100{]} \\
\texttt{descriptions} & required & 238 & 100\% & {[}98, 100{]} \\
\texttt{references} & required & 238 & 100\% & {[}98, 100{]} \\
\texttt{metrics} & optional & 238 & 100\% & {[}98, 100{]} \\
\texttt{problemTypes} & optional & 238 & 100\% & {[}98, 100{]} \\
\texttt{credits} & optional & 0 & 0.0\% & {[}0.0, 1.5{]} \\
\end{longtable}
}

The schema status column carries the comparison. \texttt{affected},
\texttt{descriptions}, and \texttt{references} are required by the CNA
published container definition {[}1{]}, so their presence says nothing
about authoring choices and we exclude them from the argument; we list
them only to show that the records are complete against the mandatory
set. The load-bearing rows are \texttt{metrics} and
\texttt{problemTypes}, which the schema leaves optional exactly as it
leaves \texttt{credits}. GitHub populates both on every record in the
sample and the third on none, with an upper confidence bound of 1.5\%.

The result holds in every cluster. Table 2 gives the per-cluster
breakdown, which is the closest thing the design supports to a statement
about persistence over time: six observation points across 26 months,
each drawn independently, all at 100\% absence.

\textbf{Table 2. GitHub-assigned records without a \texttt{credits}
container, by D1 cluster.}

{\def\LTcaptype{none} 
\begin{longtable}[]{@{}
  >{\raggedright\arraybackslash}p{(\linewidth - 8\tabcolsep) * \real{0.2000}}
  >{\raggedright\arraybackslash}p{(\linewidth - 8\tabcolsep) * \real{0.2000}}
  >{\raggedright\arraybackslash}p{(\linewidth - 8\tabcolsep) * \real{0.2000}}
  >{\raggedright\arraybackslash}p{(\linewidth - 8\tabcolsep) * \real{0.2000}}
  >{\raggedright\arraybackslash}p{(\linewidth - 8\tabcolsep) * \real{0.2000}}@{}}
\toprule\noalign{}
\begin{minipage}[b]{\linewidth}\raggedright
Cluster
\end{minipage} & \begin{minipage}[b]{\linewidth}\raggedright
Advisory dates
\end{minipage} & \begin{minipage}[b]{\linewidth}\raggedright
Records in cluster
\end{minipage} & \begin{minipage}[b]{\linewidth}\raggedright
GitHub-assigned
\end{minipage} & \begin{minipage}[b]{\linewidth}\raggedright
GitHub without credits
\end{minipage} \\
\midrule\noalign{}
\endhead
\bottomrule\noalign{}
\endlastfoot
1 & 2024-06-17 to 2024-06-28 & 27 & 19 & 19 of 19 \\
2 & 2024-12-13 to 2024-12-30 & 42 & 25 & 25 of 25 \\
3 & 2025-06-18 to 2025-06-30 & 48 & 42 & 42 of 42 \\
4 & 2025-12-18 to 2025-12-31 & 41 & 31 & 31 of 31 \\
5 & 2026-06-26 to 2026-06-30 & 71 & 53 & 53 of 53 \\
6 & 2026-08-21 to 2026-08-26 & 74 & 68 & 68 of 68 \\
\end{longtable}
}

This establishes one thing precisely: GitHub's record-authoring path is
not minimal. It emits optional structured content that requires
authoring effort. We consider in Section 5 what the pattern does and
does not license as an inference about intent.

\subsubsection{4.2 RQ2: The same omission in the OSV
export}\label{rq2-the-same-omission-in-the-osv-export}

GitHub publishes the advisory database as OSV-format files {[}25{]}. The
OSV schema defines a \texttt{credits} array with name, contact, and type
{[}24{]}, so the field exists in the target format. We retrieved the
export file for every advisory in D1.

\textbf{Table 3. \texttt{credits} in GitHub's OSV export, D1
advisories.}

{\def\LTcaptype{none} 
\begin{longtable}[]{@{}
  >{\raggedright\arraybackslash}p{(\linewidth - 6\tabcolsep) * \real{0.2500}}
  >{\raggedright\arraybackslash}p{(\linewidth - 6\tabcolsep) * \real{0.2500}}
  >{\raggedright\arraybackslash}p{(\linewidth - 6\tabcolsep) * \real{0.2500}}
  >{\raggedright\arraybackslash}p{(\linewidth - 6\tabcolsep) * \real{0.2500}}@{}}
\toprule\noalign{}
\begin{minipage}[b]{\linewidth}\raggedright
Outcome
\end{minipage} & \begin{minipage}[b]{\linewidth}\raggedright
Count
\end{minipage} & \begin{minipage}[b]{\linewidth}\raggedright
Share
\end{minipage} & \begin{minipage}[b]{\linewidth}\raggedright
95\% CI
\end{minipage} \\
\midrule\noalign{}
\endhead
\bottomrule\noalign{}
\endlastfoot
Export file retrieved & 302 of 309 & 97.7\% & {[}95, 99{]} \\
No file found in the repository & 7 of 309 & 2.3\% & {[}0.9, 4.6{]} \\
Of those retrieved, carrying \texttt{credits} & 0 of 302 & 0.0\% &
{[}0.0, 1.2{]} \\
\end{longtable}
}

Six of the seven missing files correspond to the six CVE identifiers
that CVE Services also does not serve; the seventh is a separate defect.
Both are reported in Section 4.6. The key \texttt{credits} does not
appear in the union of keys across the 302 retrieved files, which is
\texttt{affected}, \texttt{aliases}, \texttt{database\_specific},
\texttt{details}, \texttt{id}, \texttt{modified}, \texttt{published},
\texttt{references}, \texttt{schema\_version}, \texttt{severity},
\texttt{summary}, and \texttt{withdrawn}.

Every advisory in D1 has credits in GitHub's REST API, by construction
of the frame. The same advisory, exported by the same organization to a
schema that defines the field, has none. GitHub therefore serves
reporter attribution through one interface and withholds it from both
file-based formats it produces.

The longitudinal case is worth stating concretely because it is the
example the original 2023 report used. Issue \#1580 {[}23{]} names
GHSA-c653-6hhg-9x92, published January 5, 2023, and observes that the
credit shown on the advisory page is absent from the JSON. As of August
26, 2026, GitHub's REST API returns a credit for that advisory naming
user \texttt{hacdias} with type \texttt{analyst}; the OSV export file
contains no \texttt{credits} key; and the corresponding CVE record,
CVE-2023-22460, assigned by GitHub, contains no \texttt{credits}
container while carrying both \texttt{metrics} and
\texttt{problemTypes}. The specific defect reported three years and
seven months ago is unchanged, and it is present on a third surface the
original report did not examine.

\subsubsection{4.3 RQ3: Variation across assigners in the
advisory-linked
pool}\label{rq3-variation-across-assigners-in-the-advisory-linked-pool}

Table 4 disaggregates D1 by assigner. Intervals are wide for the smaller
rows and we draw no conclusions from them individually. The table lists
every assigner in D1 with at least four records and aggregates the rest.

\textbf{Table 4. Records without a \texttt{credits} container, by
assigner (D1).}

{\def\LTcaptype{none} 
\begin{longtable}[]{@{}lllll@{}}
\toprule\noalign{}
Assigner & Records & No credits & Share & 95\% CI \\
\midrule\noalign{}
\endhead
\bottomrule\noalign{}
\endlastfoot
GitHub & 238 & 238 & 100\% & {[}98, 100{]} \\
Erlang Ecosystem Foundation & 18 & 0 & 0.0\% & {[}0.0, 19{]} \\
MITRE & 9 & 9 & 100\% & {[}66, 100{]} \\
Apache & 8 & 2 & 25\% & {[}3, 65{]} \\
VulnCheck & 6 & 0 & 0.0\% & {[}0.0, 46{]} \\
VulDB & 5 & 0 & 0.0\% & {[}0.0, 52{]} \\
Red Hat & 4 & 1 & 25\% & {[}1, 81{]} \\
Nine assigners, 1 to 3 records each & 15 & 10 & 67\% & {[}38, 88{]} \\
Total & 303 & 260 & 85.8\% & {[}81, 90{]} \\
\end{longtable}
}

The aggregate 85.8\% is reported for completeness and should not be read
as an estimate of any population quantity: 78.5\% of D1 is
GitHub-assigned and GitHub is at 100\%, so the aggregate largely
restates the frame's composition. Excluding GitHub, 22 of the remaining
65 records lack credits, a rate of 33.8\%. Section 4.4 supplies the
population figure that D1 cannot.

One comparison in Table 4 does not depend on sample size. GitHub at 238
of 238 against the Erlang Ecosystem Foundation at 0 of 18 gives Fisher's
exact p = 5.3e-28. EEF records set \texttt{source.discovery} to
EXTERNAL, carry multiple typed roles, and record
\texttt{x\_generator\ \{"engine":\ "cvelib\ 1.8.0"\}}, indicating they
are produced with a freely available client for CVE Services {[}20{]}.
Two organizations authoring under the same format, one using no-cost
tooling, differ by the full range of the measure.

\subsubsection{4.4 The population census}\label{the-population-census}

D1 selects for advisories that credit someone, so it cannot say what
share of CVE records carry credits generally. D2 is a census of every
record the NVD lists as published in a two-week window, with no advisory
filter.

We enumerated every CVE identifier the NVD lists as published between
June 1 and August 14, 2026, which is 5,037 identifiers, and retrieved
the upstream record for each from CVE Services. All 5,037 were retrieved
with no failures. We exclude the 148 records in state REJECTED, since a
rejected record carries no CNA content, leaving 4,889 published records
authored by 148 distinct assigners. Of these, 2,294 carry a
\texttt{credits} container, a rate of 46.9\% with a 95\% interval of
{[}46, 48{]}.

Absence is therefore common in the population but is not the norm, and
D1's much lower aggregate is a consequence of its GitHub-heavy
composition rather than a property of the record population.

\textbf{Table 5. Records carrying a \texttt{credits} container, census
of records published June 1 to August 14, 2026 (n = 4,889 published
records). Rows shown are the assigners with the largest record counts
plus every assigner named elsewhere in the paper.}

{\def\LTcaptype{none} 
\begin{longtable}[]{@{}lllll@{}}
\toprule\noalign{}
Assigner & Records & With credits & Share & 95\% CI \\
\midrule\noalign{}
\endhead
\bottomrule\noalign{}
\endlastfoot
All published records & 4889 & 2294 & 47\% & {[}46, 48{]} \\
GitHub & 570 & 0 & 0.0\% & {[}0.0, 0.6{]} \\
VulnCheck & 473 & 434 & 92\% & {[}89, 94{]} \\
Microsoft & 438 & 0 & 0.0\% & {[}0.0, 0.8{]} \\
Linux & 416 & 0 & 0.0\% & {[}0.0, 0.9{]} \\
WPScan & 342 & 342 & 100\% & {[}99, 100{]} \\
VulDB & 291 & 290 & 100\% & {[}98, 100{]} \\
Patchstack & 207 & 207 & 100\% & {[}98, 100{]} \\
IBM & 192 & 17 & 8.9\% & {[}5.2, 14{]} \\
TuranSec & 168 & 168 & 100\% & {[}98, 100{]} \\
Red Hat & 143 & 82 & 57\% & {[}49, 66{]} \\
Wordfence & 121 & 121 & 100\% & {[}97, 100{]} \\
Apache & 96 & 60 & 62\% & {[}52, 72{]} \\
MITRE & 86 & 0 & 0.0\% & {[}0.0, 4.2{]} \\
HackerOne & 27 & 9 & 33\% & {[}17, 54{]} \\
\end{longtable}
}

Two results follow.

First, GitHub's rate is 0 of 570 in a frame that applies no advisory
filter and therefore does not condition on a reporter existing. The
upper bound is 0.6\%. This is the same behaviour D1 shows, measured
without D1's selection.

Second, the distribution across assigners is concentrated at the ends
but is not exhausted by them, and we correct an earlier characterization
of our own on this point. Restricting to the 37 assigners with at least
20 records in the window, which cover 4,321 of the 4,889 records, the
shape is as follows. Sixteen assigners are at exactly 0\%, covering
1,963 records, and no assigner in this subset lies strictly between 0\%
and 5\%; the zero group is a group of true zeros rather than of low
rates. Thirteen assigners are at 95\% or above, covering 1,346 records,
and eleven of those are at exactly 100\%. Eight assigners are
intermediate, covering 1,012 records, which is 23.4\% of the subset:
VulnCheck at 92\%, NCSC.ch at 93\%, bcorg at 91\%, Apache at 62\%, Red
Hat at 57\%, HackerOne at 33\%, IBM at 8.9\%, and CPANSec at 5\%. The
111 assigners with fewer than 20 records account for 568 records at an
aggregate 51.2\%.

The correct statement is therefore that most records come from assigners
that either always populate the field or never do, and that a
substantial minority do not behave that way. Partial adoption is real
and it is not confined to small assigners: Red Hat and Apache are
established CNAs with three-figure record counts sitting near 60\%. We
note this because our own smaller sample of Apache records, both in D1
(6 of 8) and in a 250-record pilot that preceded this census (18 of 18),
suggested a constant that the census does not support. Per-assigner
rates estimated from tens of records are unreliable, including ours.

\subsubsection{4.5 RQ4: Emitted credits do not reach the
NVD}\label{rq4-emitted-credits-do-not-reach-the-nvd}

The NVD API 2.0 response schema defines no \texttt{credits} field. Its
CVE object specifies \texttt{descriptions}, \texttt{metrics},
\texttt{weaknesses}, \texttt{references}, \texttt{configurations},
\texttt{vulnStatus}, and related keys, and none carries credit
information {[}18{]}. This is a documented structural property, not an
empirical result, and we state it as such.

We nevertheless verified that it holds in practice, because a structural
claim about a schema is not a claim about what a live service returns.
D3 takes the 43 credit-bearing records from D1 and queries the NVD for
each. All 43 were present. None contained a \texttt{credits} key, none
contained the substring ``credit'' anywhere in the returned object, and
in none did any reporter name present in the upstream record appear
anywhere in the NVD representation. The union of keys across all 43 NVD
objects is \texttt{affected}, \texttt{configurations}, \texttt{cveTags},
\texttt{descriptions}, \texttt{id}, \texttt{lastModified},
\texttt{metrics}, \texttt{published}, \texttt{references},
\texttt{sourceIdentifier}, \texttt{vulnStatus}, and \texttt{weaknesses}.
Fourteen of the 43 carry \texttt{vulnStatus} ``Analyzed'', so the
omission is not an artifact of incomplete enrichment.

As a worked example, the upstream record for CVE-2024-56512, assigned by
Apache, contains

\begin{Shaded}
\begin{Highlighting}[]
\StringTok{"credits"}\ErrorTok{:} \OtherTok{[}\FunctionTok{\{}\DataTypeTok{"lang"}\FunctionTok{:} \StringTok{"en"}\FunctionTok{,} \DataTypeTok{"type"}\FunctionTok{:} \StringTok{"finder"}\FunctionTok{,} \DataTypeTok{"value"}\FunctionTok{:} \StringTok{"Matt Gilman"}\FunctionTok{\}}\OtherTok{]}
\end{Highlighting}
\end{Shaded}

Its NVD representation has \texttt{vulnStatus} ``Analyzed'' and no
credits key, and the string ``Gilman'' does not appear in it.

The consequence is that the two hops fail for different reasons and
require different remedies. At the first hop the data is never written.
At the second it has nowhere to go.

\subsubsection{4.6 Discovery provenance and consistency
defects}\label{discovery-provenance-and-consistency-defects}

Tabulating \texttt{source.discovery} across all 303 records in D1 gives
UNKNOWN 249, EXTERNAL 28, and no value at all 26. Records that assert
UNKNOWN commonly cite, in the adjacent \texttt{source.advisory} field,
the advisory that names the reporter.

Two consistency defects surfaced as byproducts of collection and we
report them rather than treating them as attrition.

Six identifiers in D1 returned HTTP 404 from CVE Services:
CVE-2026-48717, CVE-2026-47426, CVE-2026-47424, CVE-2026-44701,
CVE-2026-54050, and CVE-2026-54049. Each appears in a published,
reviewed GitHub advisory. A published advisory referencing a CVE
identifier that the canonical service does not serve is a defect in
itself, and it affected 1.9\% of the pool.

One advisory in D1, GHSA-v9mx-4pqq-h232 for CVE-2024-21548, is served by
GitHub's REST API as a reviewed advisory published December 18, 2024,
but has no file anywhere in the advisory database repository. We
confirmed this with GitHub code search over the repository, using a
known-present advisory as a control. The bulk export that downstream
databases mirror is therefore not a complete image of what the API
serves.

\subsection{5 Discussion}\label{discussion}

\textbf{What the field data does and does not establish.} The comparison
in Table 1 is often read as evidence of deliberate selectivity, and it
does not support that reading on its own. \texttt{metrics} and
\texttt{problemTypes} correspond to severity and weakness, which
GitHub's advisory creation flow presents as ordinary steps, whereas the
same documentation introduces the credit section with the word
``optionally'' {[}22{]}. An exporter that maps the part of GitHub's
internal advisory model that is always populated, and that was never
extended to a part that is not, would produce exactly the pattern in
Table 1. Field presence alone cannot separate that explanation from a
decision about credit specifically.

What Table 1 does establish is narrower and still useful: GitHub's
record-authoring path is not minimal. It emits optional structured
content that requires authoring effort, on 100\% of records. The
omission is therefore not explained by a thin or generic export, which
is the explanation a reader would otherwise reach for first.

The question of whether the gap is known is settled by the documentary
record rather than by the field data. Issue \#1580 was opened against
the advisory database on January 7, 2023 {[}23{]}. On March 8, 2023 a
GitHub staff member replied that the credit-types work had shipped, and
that ``due to technical reasons, displaying credit information in the
JSON files would have made this epic 2-3x as much work, so we cut that
part for now,'' adding that the feedback would be used to ``more highly
prioritize getting credit information into JSON files in a future
quarter.'' The gap was identified, scoped, costed, and deferred with a
stated intent to return to it. Our measurement establishes that as of
August 26, 2026 it has not been closed, on either machine-readable
surface. We make no claim about intent to withhold credit, and none is
needed: the useful finding is that a known, acknowledged, explicitly
deferred data-quality gap has persisted for three years and seven
months.

\textbf{The cost comparison sets a bar, it does not refute the cost
claim.} We did not measure GitHub's integration cost and cannot. What we
can say is that the Erlang Ecosystem Foundation emits full typed credits
on every record it assigns using a public command line client {[}20{]},
so the format imposes no barrier and the marginal authoring cost for an
organization already holding the data is small. Integration cost inside
a large advisory pipeline is a different quantity, and it may well be
the operative constraint.

\textbf{Absence is ambiguous, and for one large assigner the ambiguity
is total.} A consumer reading a record without credits cannot
distinguish three situations: no external party was involved, a reporter
declined credit, or a reporter exists and was not recorded. For
GitHub-assigned records the third case is not merely possible, it is the
case for all 238 records in D1, each of which links an advisory that
publicly names an accepted credit. A GitHub-assigned record's empty
\texttt{credits} therefore carries no information at all. We do not
extend this to the population: D2 has no advisory linkage, so for a
non-GitHub record with no credits we cannot say which of the three
situations obtains.

\textbf{Provenance is becoming a machine-read signal.} While advisories
were read by people, the omission was cosmetic. As triage automates, the
inputs are structured records, predominantly the NVD copy and OSV
mirrors. A record that declares discovery UNKNOWN while citing a
document that answers the question is the same class of silent quality
loss the NVD literature documents for versions, products, and scores
{[}4{]}, {[}6{]}, {[}8{]}. Considerable effort has gone into provenance
for build artifacts {[}12{]}, {[}13{]}, {[}14{]}; provenance for the
reports that drive patching has received less.

\textbf{What would close the gap.} In order of how directly each
addresses what we measured. First, CNAs that already hold structured
credit data should emit it; this requires no format change on either
side and is what {[}23{]} requested. Second, the NVD schema should carry
the field, which is an additive change and would otherwise leave
diligent CNAs' credits stranded at the second hop. Third, and only
relevant once the first two hold, an explicit marker for withheld credit
would let absence be read as a decision rather than an omission. The
third measure does not address any case we measured, since an assigner
that emits no credits container will not emit a marker either.

\subsection{6 Threats to Validity}\label{threats-to-validity}

\textbf{External validity.} D1 is drawn from GitHub's advisories REST
API and is 78.5\% GitHub-assigned. Its aggregate rate is a property of
that frame and not of the CVE Program, and we do not present it as one.
D1 is additionally a cluster sample rather than a period sample: its 303
records fall in six short clusters, so within-cluster correlation is
possible and the pool cannot support a claim about any month outside
those clusters. Two things mitigate this. The clusters are six months
apart and the GitHub result is identical in all six (Table 2), and D2 is
a census with no clustering at all. The central result concerns GitHub
specifically. The cross-assigner claim rests on D2, which is a census of
a two-week window in August 2026 taken without any advisory filter. A
two-week window can still be atypical: assigners that publish in batches
are over-weighted relative to their annual share, and any assigner that
published nothing in the window is absent. The census bounds sampling
error to zero but not window effects.

\textbf{Construct validity.} We operationalize attribution as a
populated \texttt{credits} container, which measures machine-consumable
attribution rather than attribution of any kind. A record could
attribute in free text instead. We searched the description text of all
260 no-credit records in D1 against fourteen attribution patterns,
including ``reported by,'' ``discovered by,'' ``found by,'' ``credited
to,'' ``thanks to,'' and ``identified by''; none contained any of them.
We also examined every \texttt{containers.adp} structure attached to all
303 records, since Authorized Data Publishers may add content after CNA
publication; none of them carried a \texttt{credits} container.
Attribution is absent from these records in structured and unstructured
form alike.

\textbf{Internal validity.} D3 initially located only 21 of 43 records
in the NVD. Investigation showed this was an artifact of the NVD rate
limit rather than a property of the data: with request pacing at 6.5
seconds and exponential backoff, all 43 were retrieved and none carried
credits. We report this because the failure mode is easy to mistake for
a finding, and any replication of this measurement should pace requests
accordingly. The six unresolvable CVE identifiers and the one missing
OSV file are reported in Section 4.6 rather than dropped.

\textbf{Statistical treatment.} All proportions carry exact 95\%
intervals. The 0 of 43 result in D3 is presented as a schema property
with an empirical confirmation, not as an estimate; the interval on it
describes the confirmation only. We draw no conclusion from any D1 row
with fewer than 18 records. In D2 we characterize the distribution only
across assigners with at least 20 records in the window and report the
remainder as an aggregate.

\textbf{Ethics.} Reporters may request anonymity and absent credit may
reflect that. This does not explain the cases we report, because the
linked advisories name the individuals publicly and GitHub requires
credited accounts to accept the credit before display. We name only
individuals already publicly credited on a vendor advisory, and only
where doing so is necessary to show a record's contents.

\textbf{Conflict of interest.} The author is an independent security
researcher who has been credited on CVE records and therefore has a
personal interest in the practice this paper examines. One record in D3
is the author's own report. It contributes one observation to one
denominator, no result depends on it, and it is not used as an
illustrative example anywhere in the paper.

\subsection{7 Related Work}\label{related-work}

\textbf{Vulnerability data quality.} Prior work documents errors in NVD
structured fields: vulnerable version ranges {[}7{]}, inconsistency
between structured entries and their source reports {[}6{]}, and
systematic assessment with proposed corrections {[}4{]}. Croft et
al.~trace how such defects propagate into datasets used to train models
{[}5{]}. Wunder et al.~survey NVD users and report missing and
incomplete entries among their principal complaints {[}9{]}. Scoring
consistency has received separate attention {[}8{]}. The 2024 enrichment
backlog was documented largely outside the academic literature {[}10{]},
and CISA's Vulnrichment backfills enrichment from the CVE side {[}11{]}.
Attribution fields do not appear in this literature, which has
concentrated on the fields that feed scoring and matching.

\textbf{Practitioner reports of this specific gap.} The phenomenon has
been publicly reported since January 7, 2023, when an issue was opened
against GitHub's advisory database asking that credit information be
included in the OSV export {[}23{]}, following related discussion in the
VulnerableCode project. GitHub's own changelog documents the
introduction of typed credits mirroring the CVE 5.0 schema in March 2023
{[}21{]}. We therefore do not claim to have discovered the omission. We
claim to have measured it: its rate, its extent across both of GitHub's
machine-readable surfaces, its behaviour relative to other optional
fields in the same records, and its contrast with assigners operating
under the same format at no cost. D1's clusters span June 2024 to August
2026; the reported issue has been open for three years and seven months.

\textbf{Advisory pipelines.} Segal et al.~model the GitHub Security
Advisories review pipeline across more than 288,000 advisories {[}17{]};
we use the reviewed subset as a frame and measure a field that study
does not consider.

\textbf{Credit as an incentive.} The economics of disclosure treats
public credit as a central non-monetary incentive {[}15{]}, {[}16{]}.
Supply chain security has produced provenance frameworks for build
artifacts {[}12{]}, {[}13{]} and an empirical literature on SBOM
adoption {[}14{]}; these govern artifact provenance rather than report
provenance.

\subsection{8 Data Availability}\label{data-availability}

The collection scripts for all four datasets, the identifier lists, the
per-record results, and the verbatim responses from the GitHub
advisories API, the advisory database repository, CVE Services, and NVD
API 2.0 are archived at https://doi.org/10.5281/zenodo.22119153 (CC0)
and independently in the Software Heritage universal source code archive
{[}27{]} under
\texttt{swh:1:dir:e6945752ec3131c20fe7dd297bb4e29a2d98264c}. A working
mirror is at https://github.com/SyedAnas01/cve-credits-attribution. The
two archived copies are canonical and each resolves independently of the
mirror and of each other. All responses were captured on August 26,
2026.

The scripts are short and the endpoints are public, so the measurement
can be rerun cheaply, and we regard that as a feature rather than an
argument against archiving it. The archive has two uses beyond
reproduction. It fixes an August 2026 baseline against which any
subsequent change in CNA or NVD practice can be dated, which matters for
a gap whose public history is one of deferral. And the D2 census is a
complete field-presence table for every CVE record published in a fixed
window, keyed by assigner, which supports questions about optional-field
adoption other than the one we asked.

We also release \texttt{cve-credit-check}, a short script that takes a
CVE identifier and reports whether the record, the OSV export where one
exists, and the NVD copy carry attribution. It is the smallest useful
form of this measurement for an individual reporter checking their own
record.

\subsection{9 Conclusion}\label{conclusion}

GitHub collects reporter credit, requires the reporter to accept it,
displays it, serves it through its own API, and puts it in neither of
the two machine-readable formats it publishes. Across 238 CVE records
whose linked advisory publicly credits a party we found no
\texttt{credits} container, and in the same records complete coverage of
two other optional CNA-supplied fields. Across 302 OSV export files for
the same advisories we found none. A foundation using a free client for
CVE Services emits the field on every record it assigns. In a census of
every CVE record published in a two-week window, 46.9\% carry credits
and assigner behaviour is predominantly, though not entirely,
all-or-nothing. Where the field is emitted, the NVD schema has no place
to put it, and we confirmed on 43 traced records that nothing survives.
The gap was reported, acknowledged, costed, and deferred in early 2023.
This paper supplies the number, on both surfaces, three years and seven
months later.

\subsection{References}\label{references}

{[}1{]} CVE Program Quality Working Group, ``CVE JSON record format,''
schema repository, https://github.com/CVEProject/cve-schema, accessed
Aug.~26, 2026.

{[}2{]} CVE Program, ``New CVE record format enables additional data
fields at time of disclosure,'' CVE Program Blog,
https://medium.com/@cve\_program/new-cve-record-format-enables-additional-data-fields-at-time-of-disclosure-82eef1d4035e,
accessed Aug.~26, 2026.

{[}3{]} CVE Program, ``Updates for CVE record format and CVE services,''
CVE.org news, Oct.~29, 2025,
https://www.cve.org/Media/News/item/blog/2025/10/29/CVE-Record-Format-CVE-Services-Updated,
accessed Aug.~26, 2026.

{[}4{]} A. Anwar, A. Abusnaina, S. Chen, F. Li, and D. Mohaisen,
``Cleaning the NVD: comprehensive quality assessment, improvements, and
analyses,'' IEEE Transactions on Dependable and Secure Computing,
vol.~19, no. 6, pp.~4255-4269, 2022.

{[}5{]} R. Croft, M. A. Babar, and M. M. Kholoosi, ``Data quality for
software vulnerability datasets,'' in Proc. 45th International
Conference on Software Engineering (ICSE), 2023, pp.~121-133.

{[}6{]} Y. Dong, W. Guo, Y. Chen, X. Xing, Y. Zhang, and G. Wang,
``Towards the detection of inconsistencies in public security
vulnerability reports,'' in Proc. 28th USENIX Security Symposium, 2019,
pp.~869-885.

{[}7{]} V. H. Nguyen and F. Massacci, ``The (un)reliability of NVD
vulnerable versions data: an empirical experiment on Google Chrome
vulnerabilities,'' in Proc. 8th ACM Symposium on Information, Computer
and Communications Security (ASIA CCS), 2013, pp.~493-498.

{[}8{]} J. Wunder, A. Kurtz, C. Eichenmueller, F. Gassmann, and Z.
Benenson, ``Shedding light on CVSS scoring inconsistencies: a
user-centric study on evaluating widespread security vulnerabilities,''
in Proc. IEEE Symposium on Security and Privacy (SP), 2024,
pp.~1102-1121.

{[}9{]} J. Wunder, A. Corona, A. Hammer, and Z. Benenson, ``On NVD
users' attitudes, experiences, hopes, and hurdles,'' Digital Threats:
Research and Practice, Sept.~2024, doi:10.1145/3688806.

{[}10{]} VulnCheck, ``The real danger lurking in the NVD backlog,'' blog
post, 2024, https://www.vulncheck.com/blog/nvd-backlog-exploitation,
accessed Aug.~26, 2026.

{[}11{]} CISA, ``Vulnrichment,''
https://github.com/cisagov/vulnrichment, accessed Aug.~26, 2026.

{[}12{]} S. Torres-Arias, H. Afzali, T. K. Kuppusamy, R. Curtmola, and
J. Cappos, ``in-toto: providing farm-to-table guarantees for bits and
bytes,'' in Proc. 28th USENIX Security Symposium, 2019, pp.~1393-1410.

{[}13{]} Open Source Security Foundation, ``SLSA: supply-chain levels
for software artifacts,'' https://slsa.dev, accessed Aug.~26, 2026.

{[}14{]} B. Xia, T. Bi, Z. Xing, Q. Lu, and L. Zhu, ``An empirical study
on software bill of materials: where we stand and the road ahead,'' in
Proc. 45th International Conference on Software Engineering (ICSE),
2023, pp.~2630-2642.

{[}15{]} M. Finifter, D. Akhawe, and D. Wagner, ``An empirical study of
vulnerability rewards programs,'' in Proc. 22nd USENIX Security
Symposium, 2013, pp.~273-288.

{[}16{]} T. Maillart, M. Zhao, J. Grossklags, and J. Chuang, ``Given
enough eyeballs, all bugs are shallow? Revisiting Eric Raymond with bug
bounty programs,'' Journal of Cybersecurity, vol.~3, no. 2, pp.~81-90,
2017.

{[}17{]} C. Segal, P. Segal, C. E. Banjar, F. de Sant'Anna Paixao, H. S.
Borges, P. Silveira, E. S. de Almeida, J. C. S. Santos, A. Kocheturov,
G. K. Srivastava, and D. S. Menasche, ``Characterizing and modeling the
GitHub security advisories review pipeline,'' in Proc. 23rd
International Conference on Mining Software Repositories (MSR), 2026,
doi:10.1145/3793302.3793360.

{[}18{]} NIST, ``NVD vulnerability API 2.0,''
https://nvd.nist.gov/developers/vulnerabilities, accessed Aug.~26, 2026.

{[}19{]} GitHub, ``REST API endpoints for global security advisories,''
https://docs.github.com/en/rest/security-advisories/global-advisories,
accessed Aug.~26, 2026.

{[}20{]} Red Hat Product Security, ``cvelib: a library and command line
interface for the CVE Services API,''
https://github.com/RedHatProductSecurity/cvelib, accessed Aug.~26, 2026.

{[}21{]} GitHub, ``Security advisories now have multiple types of
credits,'' GitHub Changelog, Mar.~7, 2023,
https://github.blog/changelog/2023-03-07-security-advisories-now-have-multiple-types-of-credits/,
accessed Aug.~26, 2026.

{[}22{]} GitHub, ``Creating a repository security advisory,'' GitHub
Docs,
https://docs.github.com/en/code-security/security-advisories/working-with-repository-security-advisories/creating-a-repository-security-advisory,
accessed Aug.~26, 2026.

{[}23{]} Hritik14, ``Missing information in json files compared to the
advisory page,'' github/advisory-database issue \#1580, opened Jan.~7,
2023, open as of Aug.~26, 2026,
https://github.com/github/advisory-database/issues/1580. Deferral
comment by K. Catlin, Mar.~8, 2023,
https://github.com/github/advisory-database/issues/1580\#issuecomment-1460900083.

{[}24{]} Open Source Security Foundation, ``Open Source Vulnerability
(OSV) schema,'' https://github.com/ossf/osv-schema, specification at
https://ossf.github.io/osv-schema/, accessed Aug.~26, 2026.

{[}25{]} GitHub, ``GitHub Advisory Database,'' repository of OSV-format
advisory files, https://github.com/github/advisory-database, accessed
Aug.~26, 2026.

{[}26{]} CVE Program, ``CVE Numbering Authorities (CNAs) partner list,''
https://github.com/CVEProject/cve-website/blob/dev/src/assets/data/CNAsList.json,
543 entries as of Aug.~26, 2026.

{[}27{]} Software Heritage, universal source code archive, INRIA and
UNESCO. Replication package archived as
\texttt{swh:1:dir:e6945752ec3131c20fe7dd297bb4e29a2d98264c}, revision
\texttt{swh:1:rev:e9376be395564a87645646a8bf668c7d8ea122ce}, deposited
Aug.~26, 2026, https://archive.softwareheritage.org.

\end{document}